\documentclass[conference]{IEEEtran}
\IEEEoverridecommandlockouts
\usepackage{cite}
\usepackage{amsmath,amssymb,amsfonts}
\usepackage{graphicx}
\usepackage{algorithm}
\usepackage{algpseudocode}
\usepackage{textcomp}
\usepackage{xcolor}
\usepackage{fancyhdr}

\fancypagestyle{icecsheader}{
    \fancyhf{}
    \fancyhead[C]{%
        \fontsize{8}{9}\selectfont
        2026 IEEE 33rd International Conference on Electronics, Circuits and Systems (ICECS)}

}

\fancypagestyle{icecsfirstpage}{
    \fancyhf{}
    \fancyhead[C]{%
        \fontsize{8}{9}\selectfont
        2026 IEEE 33rd International Conference on Electronics, Circuits and Systems (ICECS)}
    \fancyfoot[L]{%
        \fontsize{8}{9}\selectfont
        979-8-3195-1905-4/26/\$31.00 \textcopyright{}2026 IEEE}

}

\renewcommand{\baselinestretch}{0.922}

\newcommand\freefootnote[1]{%
  \let\thefootnote\relax%
  \footnotetext{#1}%
  \let\thefootnote\svthefootnote%
}
\def\BibTeX{{\rm B\kern-.05em{\sc i\kern-.025em b}\kern-.08em
    T\kern-.1667em\lower.7ex\hbox{E}\kern-.125emX}}
\begin{document}

\title{A Pipelined FPGA Architecture for Banded Sparse Matrix–Dense Matrix Multiplication in Longformer
}

\author{
    \IEEEauthorblockN{Phillip Pramberger\textsuperscript{†}, Athanasios Tziouvaras\textsuperscript{\ddag}, Shreejith Shanker\textsuperscript{†}, George Floros\textsuperscript{†}}
    \IEEEauthorblockA{
    \textsuperscript{†}Department of Electronic and Electrical Engineering, Trinity College Dublin, Ireland\\
    \textsuperscript{\ddag}Department of Electrical and Computer Engineering, 
University of Thessaly, Greece \\
%\{pramberp, shankers, florosg\}@tcd.ie, attziouv@e-ce.uth.gr
pramberp@tcd.ie, attziouv@e-ce.uth.gr, shankers@tcd.ie, florosg@tcd.ie
}}

\maketitle
% \IEEEpubid{%
% \makebox[\columnwidth][l]{%
% \fontsize{8}{9}\selectfont
% 979-8-3195-1905-4/26/\$31.00~\textcopyright~2026 IEEE}
% \hspace{\columnsep}
% \makebox[\columnwidth]{}}
% \thispagestyle{icecsfirstpage}

\begin{abstract}
Sparse attention mechanisms have become increasingly important for transformer models processing long input sequences due to their lower computational and memory complexity compared to full self-attention. Longformer achieves this through a sliding-window attention mechanism that produces a structured banded sparse attention matrix. However, existing sparse transformer accelerators primarily target attention generation or unstructured sparsity, leaving sparse matrix--dense matrix multiplication (SpMM) for structured sparse attention largely unexplored.
This paper presents a pipelined FPGA architecture for accelerating banded SpMM in Longformer. The proposed design exploits the predictable sparsity pattern of Longformer’s attention matrix through a custom row-wise storage scheme with implicit indexing, eliminating the overhead of conventional sparse matrix formats while enabling regular memory accesses. The architecture employs parallel processing elements, pipelined adder trees, and a dual-path computation strategy to maximize throughput and hardware utilization.
Implemented in Verilog and evaluated on an RFSoC platform using Vivado 2024.2, the accelerator sustains one complete dot-product result per clock cycle after an initial latency of 11 cycles while maintaining power consumption below 2.9 W. Operating at 100 MHz, the design achieves over 100 million dot-product outputs per second, demonstrating the effectiveness of directly exploiting structured sparsity for sparse transformer acceleration.
\end{abstract}


\section{Introduction}
Transformer-based models have become the dominant architecture in modern machine learning due to their strong performance across natural language processing, computer vision, and sequence modelling tasks \cite{khan2022transformers}. \freefootnote{*This work was funded under the COIN-3D project, which has received funding from the European Union’s Horizon Europe research and innovation program under grant agreement No. 101159667.}However, the self-attention mechanism at the core of transformers introduces significant computational and memory demands, particularly for long input sequences, as its complexity scales quadratically with sequence length. To address this limitation, sparse attention mechanisms have been proposed to reduce unnecessary computations by restricting attention to selected regions of the input \cite{farina2024sparsity}. Although this significantly improves scalability, efficiently accelerating sparse matrix operations on hardware remains challenging due to irregular memory accesses, load imbalance, and the overhead associated with conventional sparse matrix representations. As a result, efficient hardware support for sparse matrix–dense matrix multiplication (SpMM) has become an important area of research for scalable transformer acceleration \cite{liu2024fpga}.

Recent works have explored a variety of hardware accelerators for sparse transformers and sparse matrix operations. Architectures such as SWAT~\cite{SWAT} and SALO~\cite{SALO} accelerate sparse attention mechanisms through optimized dataflow scheduling, kernel fusion, and pipelined attention computation. Similarly, SpAtten~\cite{wang2021spatten} reduces computational overhead by dynamically pruning tokens and attention heads during runtime. While these approaches successfully improve the efficiency of sparse transformers, they primarily focus on attention generation and scheduling rather than SpMM itself.

% Alongside transformer accelerators, extensive work has also been carried out on general SpMM and SpMV acceleration. Architectures such as Serpens~\cite{song2023serpens}, Graph-OPU~\cite{chen2022graphopu}, and Leda~\cite{yi2024leda} employ pipelined processing elements, optimized sparse storage formats, and high-bandwidth memory to improve throughput for unstructured sparse matrices. However, these accelerators are largely designed for dynamic or irregular sparsity patterns, requiring complex indexing and irregular memory-access mechanisms. In contrast, this work specifically targets the structured banded sparsity produced by Longformer's sliding-window attention mechanism. By exploiting the predictable structure of the attention matrix, the proposed FPGA architecture enables regularized memory access and deterministic computation scheduling while avoiding the overhead associated with conventional sparse matrix representations.

To address these limitations, this paper proposes a pipelined FPGA architecture for accelerating banded sparse matrix--dense matrix multiplication in Longformer. The main contributions are as follows. \textit{First}, a pipelined FPGA architecture tailored to the structured banded sparsity pattern of Longformer is proposed to accelerate SpMM. \textit{Second}, a custom row-wise sparse storage scheme with implicit indexing is introduced to regularize memory access patterns while reducing the overhead associated with conventional sparse matrix formats. \textit{Third}, the architecture employs a dual-path computation strategy together with parallel processing elements and pipelined adder trees to sustain one complete dot-product output per clock cycle after initial latency. Finally, the proposed design is implemented in Verilog and evaluated on an RFSoC platform using Vivado, demonstrating high throughput, low latency, and power consumption below 2.9 W.

The remainder of this paper is structured as follows. Section \ref{sec:Back} provides background on Longformer and sparse attention mechanisms, with a particular focus on sparsification and structured banded attention matrices. Section \ref{sec:Prop} presents the proposed FPGA architecture, including the sparse storage scheme, computation pipeline, and overall system design. Section \ref{sec:Exp} discusses the implementation details and experimental evaluation carried out on an RFSoC platform using Vivado. Finally, Section \ref{sec:Concl} concludes the paper.

\begin{figure*}[!hbt]
    \begin{center}
    
\includegraphics[width=1.95\columnwidth]{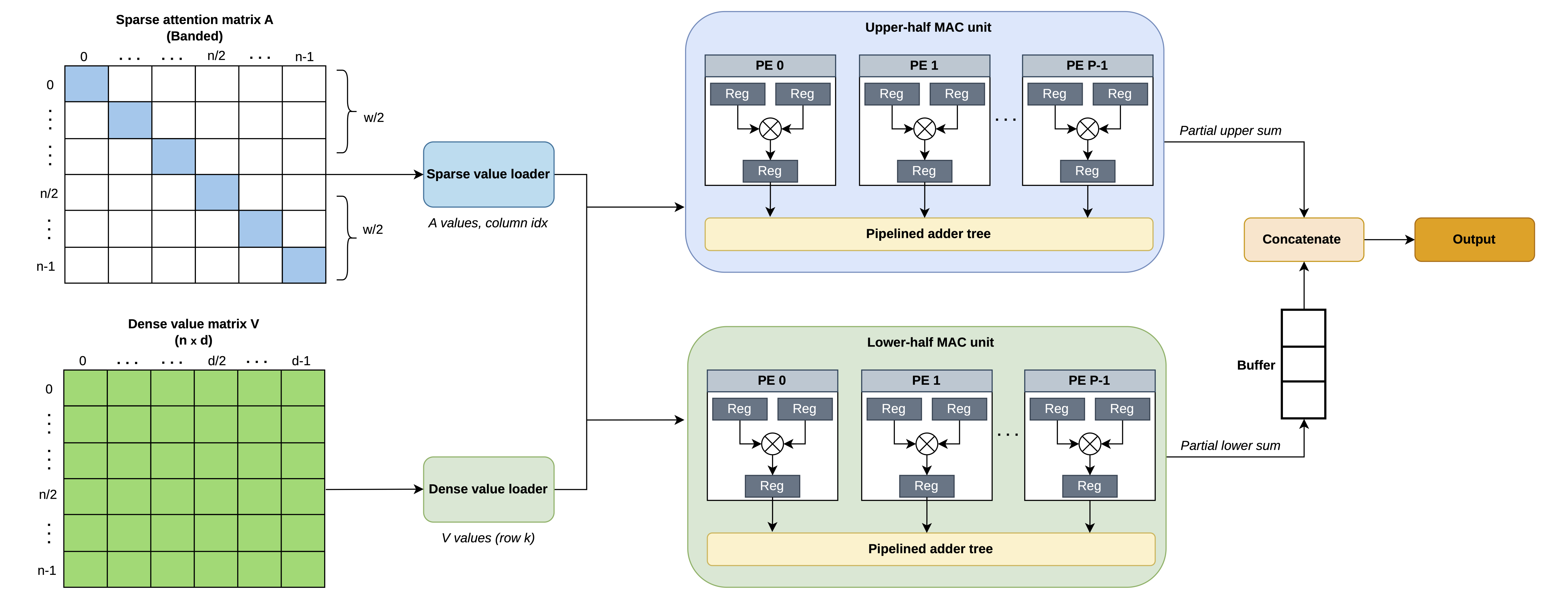}
    \caption{Proposed Pipelined FPGA Architecture for Longformer Banded SpMM.}
    \label{fig:architecture}
     \end{center}
\end{figure*}

\section{Background and Motivation}\label{sec:Back}

\subsection{Transformer Self-Attention}

Transformer models rely on the self-attention mechanism introduced in \textit{Attention Is All You Need}~\cite{vaswani2017attention}. Given the query, key, and value matrices $Q,K,V \in \mathbb{R}^{n \times d}$, the attention operation is defined as

\begin{equation}
\mathrm{Attention}(Q,K,V)=
\mathrm{softmax}\left(
\frac{QK^T}{\sqrt{d_k}}
\right)V,
\label{eq:attention}
\end{equation}
where $n$ denotes the sequence length and $d_k$ represents the dimensionality of the key vectors. The matrix

\begin{equation}
A=
\mathrm{softmax}
\left(
\frac{QK^T}{\sqrt{d_k}}
\right)
\end{equation}
contains the attention weights and has dimensions $n \times n$. Since every token attends to every other token, the computational and memory complexity of full self-attention scales quadratically with sequence length~\cite{vaswani2017attention}, i.e., $\mathcal{O}(n^2 d)$. This quadratic scaling becomes prohibitively expensive for long-sequence applications due to the large number of matrix operations and memory accesses required.

\subsection{Longformer and Sparse Attention}

Longformer~\cite{beltagy2020longformer} addresses the scalability limitations of full self-attention by employing a sliding-window sparse attention mechanism. Instead of attending to all tokens in a sequence, each token attends only to neighboring tokens within a fixed local window of width $w$. Consequently, the attention matrix becomes sparse and banded, where non-zero entries satisfy

\begin{equation}
|i-j| \leq \frac{w}{2}
\end{equation}
for row index $i$ and column index $j$. The resulting sparse attention matrix therefore contains non-zero values concentrated around the main diagonal, producing a structured banded sparsity pattern.

By restricting attention to local neighborhoods, Longformer reduces the complexity of attention computation from quadratic to approximately $\mathcal{O}(nwd)$,
where $w \ll n$~\cite{beltagy2020longformer}. This significantly improves scalability for long-sequence transformer workloads while preserving model performance.

\subsection{Sparse Matrix--Dense Matrix Multiplication}

Following sparsification, the dominant computation becomes SpMM:

\begin{equation}
Y = AV,
\end{equation}
where
$A \in \mathbb{R}^{n \times n}$ is the sparse banded attention matrix and $V \in \mathbb{R}^{n \times d}$ is the dense value matrix. For a banded sparse matrix, each output element can be expressed as:

\begin{equation}
Y_{ij}=
\sum_{k \in \mathcal{N}(i)}
A_{ik}V_{kj},
\end{equation}
where $\mathcal{N}(i)$ denotes the non-zero neighborhood defined by the sliding-window attention mechanism.

Efficiently accelerating SpMM remains challenging due to irregular memory accesses, sparse indexing overhead, and load imbalance~\cite{liu2024fpga}. However, unlike unstructured sparse matrices, the predictable banded structure produced by Longformer enables regularized memory access patterns and deterministic computation scheduling, making it particularly suitable for FPGA-based acceleration.

\section{Architecture Design} \label{sec:Prop}

\subsection{Architecture Overview}

The proposed accelerator is designed to efficiently compute SpMM for the structured banded attention matrices produced by Longformer. Given a sparse attention matrix $A \in \mathbb{R}^{n \times n}$ and dense value matrix $V \in \mathbb{R}^{n \times d}$, the architecture computes the output matrix $Y = AV$ by exploiting the predictable sparsity pattern of $A$. As illustrated in Fig.~\ref{fig:architecture}, the design consists of sparse and dense value loaders, parallel multiply--accumulate (MAC) units, pipelined processing elements (PEs), and hierarchical adder trees.

The proposed architecture follows a dual-path computation strategy inspired by parallel algorithms for banded matrix multiplication \cite{fox2014parallel}. The sparse attention matrix is partitioned into upper and lower band components that are processed concurrently to maximize parallelism and sustain pipeline utilization, as shown in Alg. \ref{alg:dualpath}.

\begin{algorithm}[!t]
\caption{Dual-Path Banded Sparse Matrix--Dense Matrix Multiplication}
\label{alg:dualpath}
\begin{algorithmic}[1]

\For{$i = 0$ to $n-1$}
 \For{$j = 0$ to $d-1$}
    \State $Y^{U}_{ij} \gets 0$
    \State $Y^{L}_{ij} \gets 0$

    \Statex \textit{// Compute upper and lower band contributions in parallel}

    \State \textbf{parallel}

    \For{$k = \max(0,i-\frac{w}{2})$ to $i$}
        \State $Y^{U}_{ij} \gets Y^{U}_{ij} + A_{ik}V_{kj}$
    \EndFor

    \For{$k = i+1$ to $\min(n-1,i+\frac{w}{2})$}
        \State $Y^{L}_{ij} \gets Y^{L}_{ij} + A_{ik}V_{kj}$
    \EndFor

    \State \textbf{end parallel}

    \Statex \textit{// Combine partial sums}

    \State $Y_{ij} \gets Y^{U}_{ij} + Y^{L}_{ij}$
\EndFor
\EndFor

\end{algorithmic}
\end{algorithm}

The upper and lower partial sums are computed simultaneously using independent MAC units composed of $P$ parallel PEs and pipelined reduction trees. This allows multiple sparse matrix products to be computed concurrently before reduction and accumulation to produce the final output element. By exploiting the deterministic sparsity pattern of Longformer, the architecture enables fixed scheduling, regular memory accesses, and sustained throughput of one output per clock cycle after the initial pipeline latency, while avoiding the control overhead typically associated with unstructured sparse matrix accelerators.

\subsection{Structured Sparse Storage Format}

The proposed architecture exploits the structured banded sparsity pattern produced by Longformer’s sliding-window attention mechanism. Since non-zero values are concentrated around the main diagonal, only elements satisfying

\begin{equation}
|i-j| \leq \frac{w}{2}
\end{equation}
are stored and processed, where $w$ denotes the attention window size. Unlike conventional sparse matrix formats such as Compressed Sparse Row (CSR) or Coordinate List (COO), the proposed design employs a regular row-wise storage scheme tailored to the deterministic structure of the attention matrix.

For each row, non-zero values are stored sequentially according to their diagonal offset, while rows near matrix boundaries are zero-padded to maintain a constant window width. Since the position of each value within the fixed-width band implicitly determines its column offset, explicit column indices are not required. This enables implicit indexing and regular memory access patterns, eliminating complex pointer traversal and reducing control overhead compared to CSR- and COO-based representations. The sparse matrix is further partitioned into upper and lower band components that are processed independently by parallel MAC units to maximize pipeline utilization and computation throughput.

\subsection{Processing Elements and Computation Pipeline}

The proposed architecture employs parallel PEs to compute sparse matrix--dense matrix products concurrently. Each PE performs MAC operations between non-zero sparse matrix values and corresponding dense matrix elements,

\begin{equation}
P_k = A_{ik}V_{kj}.
\end{equation}

The resulting partial products are reduced through a hierarchical pipelined adder tree to generate partial sums for each output element. To exploit the structured banded sparsity pattern of Longformer, the computation is divided into upper and lower band components that are processed simultaneously by independent MAC units. The upper-half MAC unit computes contributions from the main diagonal and entries below the diagonal, while the lower-half MAC unit processes contributions from entries above the diagonal.

Both computation paths operate concurrently and produce partial sums that are accumulated in a final reduction stage to obtain the output value $Y_{ij}$. Pipeline registers are inserted between reduction stages to sustain continuous dataflow and maximize operating frequency. After the initial pipeline latency, the architecture sustains one complete dot-product output per clock cycle through continuous pipelined execution.

\subsection{Dataflow and Memory Organization}

Efficient memory access is critical for sustaining high-throughput SpMM computation. The proposed architecture employs separate sparse and dense value loaders to stream data into the computation pipeline. Sparse matrix values are fetched sequentially according to their banded row-wise organization, while the corresponding dense matrix rows are streamed and broadcast to the parallel processing elements.

Due to the deterministic sparsity pattern of Longformer, memory accesses remain highly regular and predictable. This eliminates the irregular pointer traversal and indirect addressing commonly associated with conventional sparse matrix formats. Rows near matrix boundaries are zero-padded to preserve a constant access pattern and maintain pipeline continuity across all computations.

The sparse matrix is partitioned into upper and lower band components that are streamed independently into dedicated MAC units. Partial sums generated by both computation paths are buffered and accumulated in the final reduction stage. By maintaining continuous streaming and pipelined dataflow throughout the architecture, the design sustains one complete output result per clock cycle after an initial pipeline latency of 11 cycles. This organization minimizes control overhead and enables sustained pipeline utilization throughout the~computation.

\section{Implementation and Evaluation} \label{sec:Exp}

%- Experimental set up (FPGA etc)\\
%- A big table with all the results with different optimizations\\
%- Discussion of the table\\
%- Overview of the achievements (GPU discussion)
\subsection{Implementation Results}
The proposed accelerator was implemented in Verilog and synthesized using Vivado 2024.2 targeting the RFSoC~4$\times$2 platform operating at 100~MHz. Functional simulation verified correct operation and confirmed that, after an initial pipeline latency of 11 clock cycles, the architecture sustains a throughput of one complete dot-product result per clock cycle. At the target clock frequency, this corresponds to a peak throughput exceeding 100 million dot-product outputs per second.

Three implementation runs were conducted using the default, area-optimized, and power-optimized Vivado implementation strategies. Table~\ref{tab:impl_results} summarizes the resulting resource utilization, power consumption, and timing characteristics. Across all implementation runs, LUT utilization remained between 65k and 70k units, while DSP utilization remained constant at 2052 slices, corresponding to approximately 48\% of the available DSP resources on the target device. FF utilization remained stable at approximately 91k registers, indicating that the optimization strategies had minimal impact on the architecture's pipelined datapath. Power consumption remained below 2.9~W in all cases.

Timing analysis further confirmed successful implementation, with positive Worst Negative Slack (WNS) and Worst Hold Slack (WHS) values across all implementation strategies, indicating the absence of timing violations. Although the area-optimized implementation reduced LUT utilization by approximately 6\%, the resulting gains were relatively modest and accompanied by a slight reduction in timing margin. Similarly, the power-oriented implementation produced only negligible changes in power consumption while also reducing timing slack. Consequently, the default implementation provides the most balanced trade-off between resource utilization, power consumption, and timing performance.

\begin{table}[h]
\centering
\caption{Implementation Results of the Proposed Accelerator}
\label{tab:impl_results}
\resizebox{\columnwidth}{!}{%
\begin{tabular}{|l|c|c|c|c|c|}
\hline
\textbf{Implementation} &
\textbf{LUTs} &
\textbf{DSPs} &
\textbf{Power (W)} &
\textbf{WNS (ns)} &
\textbf{WHS (ns)} \\
\hline
Default & 70,177 & 2,052 & 2.879 & 0.755 & 0.010 \\
Area Optimized & 65,923 & 2,052 & 2.883 & 0.729 & 0.010 \\
Power Optimized & 70,183 & 2,052 & 2.887 & 0.658 & 0.011 \\
\hline
\end{tabular}
}
\end{table}

\subsection{Comparison with a Dense Baseline}

To evaluate the benefits of exploiting Longformer's structured sparsity, a comparative baseline architecture was constructed in which the operation was treated as a conventional multiplication between two dense $2048 \times 2048$ matrices. Unlike the proposed design, which leverages the fixed-width banded structure of the attention matrix to eliminate unnecessary computations and indexing overhead, the dense baseline performs all multiplications regardless of sparsity.

Despite using 16-bit signed fixed-point arithmetic rather than the 32-bit fixed-point representation used by the proposed accelerator, the dense implementation requires more than 370,000 LUTs, corresponding to more than five times as many LUTs as the proposed accelerator, as shown in Table \ref{tab:dense_compare}. Although DSP utilization was reduced to 896 slices, the substantial increase in logic utilization exceeded the capacity of many FPGA devices. Furthermore, the power analysis estimated a total on-chip power consumption exceeding 3~kW, resulting in severe thermal violations and rendering the design impractical for deployment on the target platform.

These results demonstrate that directly exploiting the banded sparsity pattern produced by Longformer significantly improves hardware efficiency. By eliminating redundant computations associated with zero-valued entries and avoiding the overhead of conventional sparse matrix indexing schemes, the proposed architecture achieves a practical implementation with modest resource utilization, low power consumption, and successful timing closure.

\begin{table}[h]
\centering
\caption{Comparison Between the Proposed Sparse Accelerator and a Dense Baseline}
\label{tab:dense_compare}
\begin{tabular}{|l||c|c|}
\hline
\textbf{Metric} &
\textbf{Proposed Accelerator} &
\textbf{Dense Baseline} \\
\hline \hline
Matrix Type & Banded Sparse & Dense \\
Precision & 32-bit Fixed Point & 16-bit Fixed Point \\
LUTs & 70,177 & $>$370,000 \\
DSPs & 2,052 & 896 \\
Power Consumption & 2.879 W & $>$ 3 kW (estimated) \\
\hline
\end{tabular}
\end{table}

\subsection{Discussion}
By exploiting Longformer's banded sparsity pattern, the proposed accelerator eliminates the majority of multiplications involving zero-valued elements while maintaining regular memory accesses, resulting in a computational complexity of $O(nwd)$, where $n$ denotes the sequence length, $w$ the attention window size, and $d$ the number of columns in the dense matrix.

In contrast, dense GPU approaches such as CUTLASS~\cite{nvidia_cutlass_gemm} treat the operation as a GEMM problem and do not explicitly exploit the fixed sparsity structure, leading to a workload that scales approximately as $O(n^2d)$. Sparse libraries such as cuSPARSE~\cite{nvidia_cusparse} can reduce this complexity through CSR- or COO-based representations, but incur indexing overhead and may introduce irregular memory accesses~\cite{bell2009spmv}. By leveraging the deterministic structure of Longformer's attention matrix, the proposed architecture eliminates explicit indexing, enables predictable dataflow, and reduces memory overhead, making it well suited for accelerating structured sparse  workloads.

\section{Conclusions} \label{sec:Concl}

This paper presented a pipelined FPGA architecture for accelerating banded sparse matrix--dense matrix multiplication in Longformer by exploiting the structured sparsity produced by its sliding-window attention mechanism. The proposed design employs a custom storage format with implicit indexing, pipelined processing elements, and a dual-path computation strategy to achieve efficient parallel execution and regular dataflow. Implemented in Verilog and evaluated on an RFSoC platform, the accelerator sustains one complete dot-product output per clock cycle after an initial latency of 11 cycles while operating at 100~MHz with power consumption below 2.9~W. Using approximately 70k LUTs and 2052 DSP slices, the design achieved successful timing closure and demonstrated the benefits of directly exploiting structured sparsity for FPGA-based sparse transformer acceleration.

\bibliographystyle{IEEEtran}
\bibliography{refs}
\end{document}